\documentclass[final]{eps}

\usepackage{graphicx}
\usepackage{times}

\volumenumber{}
\publishyear{}
\frompage{\pageref{firstpage}}
\topage{\pageref{finalpage}}

\shorttitle{P. Shalima, J. Murthy and R. Gupta: Dust properties from UV halo around Spica}

\title{Dust properties from GALEX observations of a UV halo around Spica}
\author{P. Shalima$^1$, Jayant Murthy$^2$, and Ranjan Gupta$^1$}
\affiliation{$^1$Inter-University Center for Astronomy and Astrophysics, Pune\\
             $^2$Indian Institute of Astrophysics, Bangalore}
\received{xxxx xx, 2003}\revised{xxxx xx, 2003}\accepted{xxxx xx, 2003}
\abstract{GALEX has detected ultraviolet halos extending as far as 5$^{\circ}$ around four bright stars (Murthy et al. (2011)). These halos are produced by scattering of starlight by dust grains in thin foreground clouds that are not physically associated with the star.  Assuming a simple model consisting of a single layer of dust in front of the star, Murthy et al.(2011) have been able to model these halo intensities and constrain the value of the phase function asymmetry factor $g$ of the scattering grains in the FUV and NUV. However due to the uncertainty in the dust geometry they could not constrain the albedo. 
In this work we have tried to constrain the optical constants and dust geometry by modeling the
 UV halo of Spica. Since the halo emission is not symmetric, we have modeled the Northern and Southern parts of the halo separately.
To the North of Spica, the best-fit albedo is 0.26$\pm$0.1 and $g$ is 0.58$\pm$0.11 in the FUV at the 90\% confidence level. The corresponding limits on the distance and optical depth ($\tau$) of the dust sheet is 3.65$\pm$1.05 pc and 0.047$\pm$0.006 respectively. 
 However, owing to a complicated dust distribution to the South of Spica, we were unable to uniquely constrain the dust parameters in that region. Nevertheless, by assuming the optical constants of the Northern region and assuming a denser medium, we were able to constrain the distance of the dust to 9.5$\pm$1.5 pc and the corresponding $\tau$ to 0.04$\pm$0.01.           
}

\begin{document}
\label{firstpage}
\maketitle
\copyrighttext{}

\section{Introduction}
Haloes around stars are a consequence of forward scattering of starlight from a thin layer of dust between us and the star. Hence they provide a useful tool to determine the properties of dust grains in the ISM. There have been observations of X-ray halos around stars and recently Murthy et al. (2011) have discovered UV halos around four bright stars including Spica.
By modeling these UV halos, Murthy et al. (2011) have placed limits of 0.58$\pm$0.12  and 0.72$\pm$0.02 on the phase function asymmetry factor $g$ in the FUV and NUV respectively. However, due to an uncertain geometry they were unable to constrain the albedo and the dust distribution towards Spica.

\begin{figure*}[ht]
\centerline{\includegraphics[width=10cm,clip]{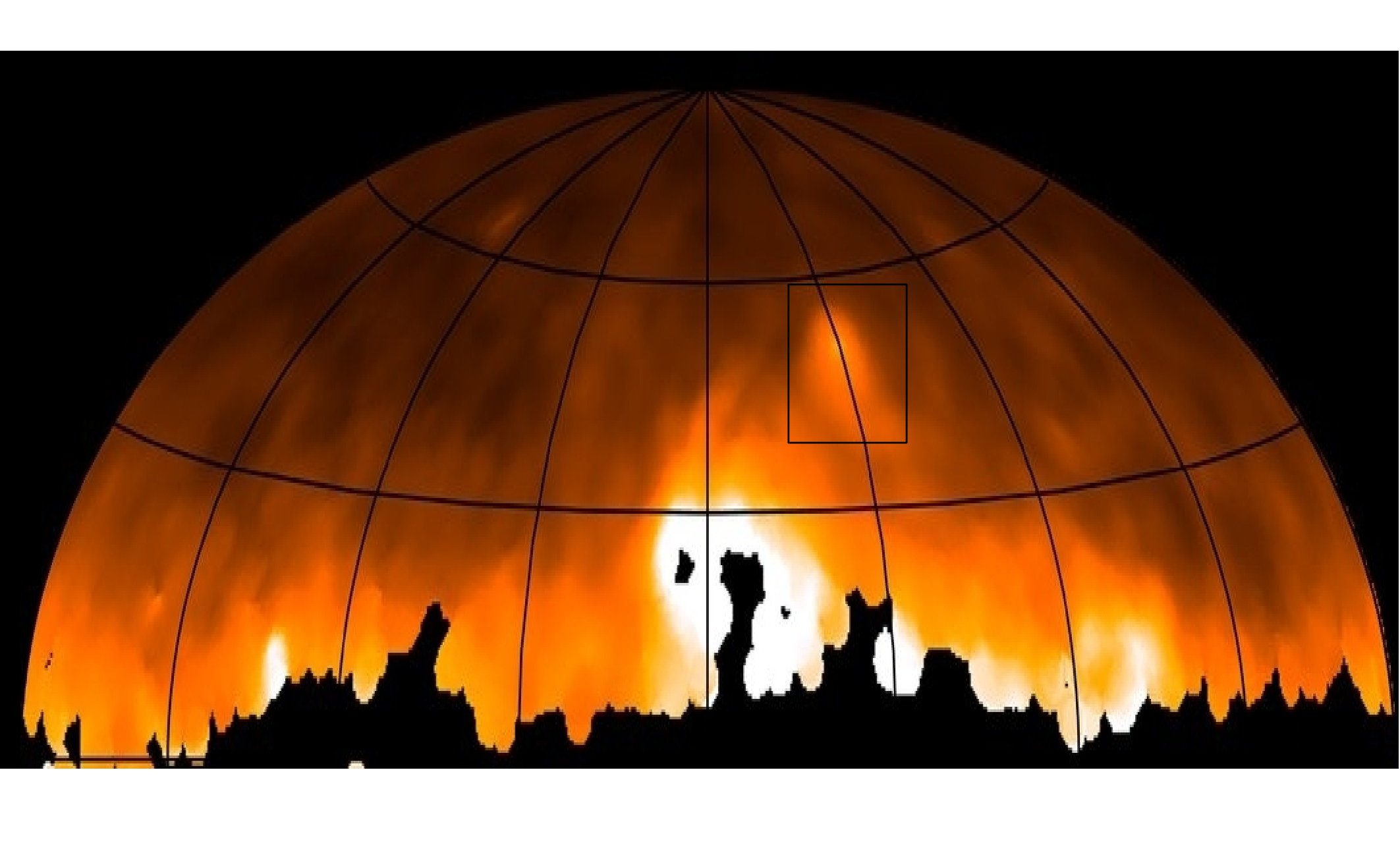}} 
\caption{GALEX FUV image of the diffuse sky with the halo around Spica shown within the square in the Northern galactic hemisphere (Murthy et al. 2011).}
\label{image}
\end{figure*}
Spica (l=316.11$^\circ$, b= 50.84$^\circ$) also known as Alpha Virginis is a B1 III-IV star in a binary system at a distance of 80pc from us, that hosts an HII region around it. 
The gas column density in front of the star is low (N$_{H}$ $\sim$ 10$^{19}$ cm$^{-2}$).
Despite the low column density, a bright UV halo has been observed around it.
But the presence of the interaction zone between the local and Loop I bubbles to the South of Spica (Park et al. 2010) makes the dust geometry complicated and the halo asymmetric. The region has a low NaI/CaII region suggesting that it has been exposed to shock waves (Centurion \& Vladilo 1991). 
There is also evidence of a significant enhancement in the abundance of very small grains (by a factor of 2 to 3) in the region (Zagury et al. 1998) which has been attributed to the destruction of large grains in shocks.

Here we have attempted to constrain the optical constants of dust grains and their distribution toward Spica by taking into account the asymmetry in the dust distribution to the North and South of Spica.

\section{Observations}
We have used GALEX observations of the Spica halo, which are part of the All-Sly Imaging survey with exposure times of $\sim$100 s (see Fig \ref{image}). The data analysis has been described in Murthy \& Henry (2011). The nearest observation was 2$^{\circ}$ away from the star since GALEX did not observe close to very bright stars.
\section{Model}

We have modeled the halo intensities by considering the dust to be in the form of two layers in front of Spica and assuming single scattering due to the low column densities involved.
From the 60/100 micron ratios, Murthy et al. (2011) have derived a distance of 3 pc for the scattering layer. 
Rather than using a fixed distance we have allowed the scattering layer distance to vary between 1-10 pc from the star. We have placed a 2nd sheet at 40 pc which contributes ~20-30\% of the scattered intensity and only provides an offset to the intensity. Varying this sheet anywhere between 30 and 70 pc from the star does not cause a change of more than 25\% in the model intensities. 
 
Even though column density measurements are available for Spica, that is applicable only for the region right in front of Spica and not necessarily for the UV halo for which we have observations 2 degrees away from the star, corresponding to a distance of 2.8 pc from it. 
In fact extinction measurements of stars in the region show that there is enough material at a distance less than $\sim$90 pc with an optical depth of 0.09 (D. O. Jones, A. A. West \& J. B. Foster, 2011).
Hence the optical depth of the scattering layer is varied between 0 and 0.1, but keeping the total optical depth fixed at 0.1. 

The stellar flux has been calculated using the Kurucz (1979) model spectra convolved with the instrument calibration. For the angular dependence of scattering, we have made use of the Henyey-Greenstein scattering function (Henyey \& Greenstein 1941). The optical constants, albedo and $g$ are varied between 0.1 to 1.0 and 0.0 to 1.0 respectively. 

\section{Results} 
\subsection{Northern region}
For the Northern region we have compared our model with the observations 
and derived 
confidence levels for the two optical constants albedo and $g$ as well as the dust distribution and optical depth of the scattering layer. The 68\% and 90\% confidence levels are obtained from the $\chi^{2}$ values for each set of parameters using the following equation of Lampton , Margon \& Bowyer (1976);
$$
S_{L} = S_{min} + \chi^{2}_{p}(\alpha)
$$

where, $S_{L}$ is the contour corresponding to significance $\alpha$, $S_{min}$ is the minimum value of $\chi^{2}$ and $\chi^{2}_{p}(\alpha)$ is the tabulated value of $\chi^{2}$ for $p$ parameters and significance $\alpha$.

\begin{figure}[h]
\centerline{\includegraphics[width=8cm,clip]{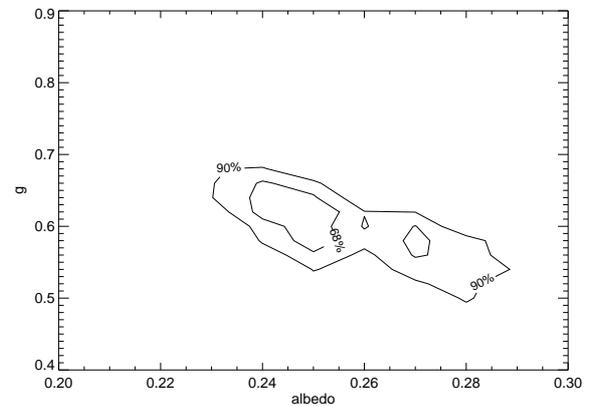}} 
\caption{68\% and 90\% contour plot of the albedo and $g$ of the dust cloud in the FUV for the Northern regions.}
\end{figure}
\begin{figure}[h]
\centerline{\includegraphics[width=8cm,clip]{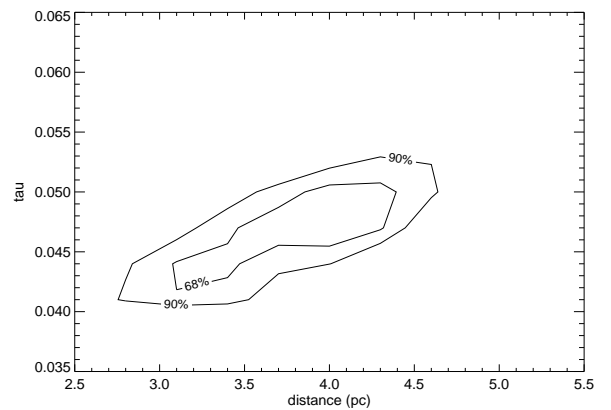}} 
\caption{68\% and 90\% contour plot of the distance and optical depth of the dust cloud in the FUV for the Northern regions.}
\end{figure}
The best-fit values are 0.26$\pm$0.1 for the albedo and 0.58$\pm$0.11 for $g$ in the FUV at 90\% confidence. The error bars for the albedo have been increased to 0.1 to account for uncertainties in the model geometry. The corresponding limits on the distance and $\tau$ are 3.65$\pm$0.05 pc and 0.047$\pm$0.006 respectively (see Figures 1. and 2.). 
 
%\subsubsection{Subsubsection}\strut
We have then compared the best-fit model intensities with the observed values (see Figure 3). The values are well correlated with a co-efficient of 0.91.   
\begin{figure}[h]
\centerline{\includegraphics[width=8cm,clip]{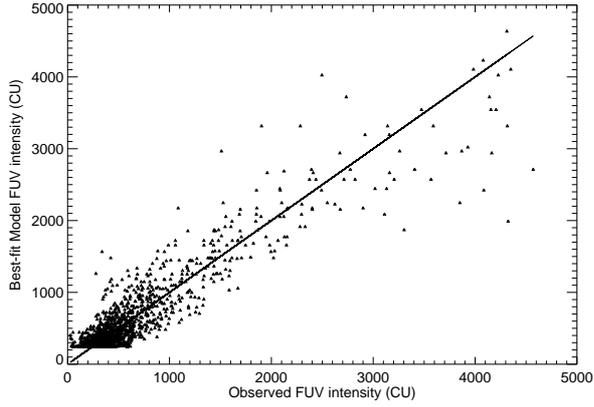}} 
\caption{Comparison of Observed and Model intensities in the FUV towards the North of Spica.}
\end{figure}
\begin{figure}[h]
\centerline{\includegraphics[width=8cm,clip]{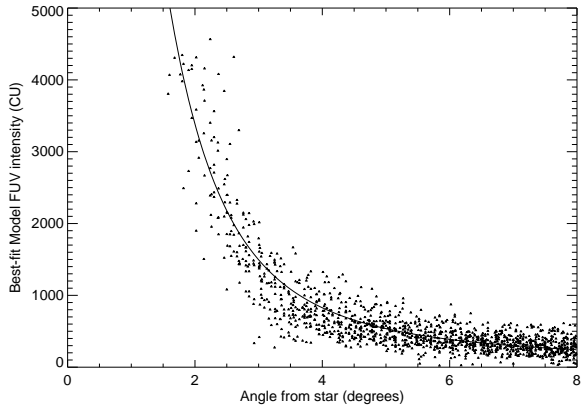}} 
\caption{Angular dependence of the observed and model FUV intensity of the halo towards the North of Spica.}
\end{figure}

The angular dependence of the observed (solid line) and model ('+' signs) intensities in the FUV is shown in Figure 4. It is evident that the model gives a reasonably good fit to the observed intensity profile for the Northern parts of the halo.
\subsection{Southern region}

\begin{figure}[h]
\centerline{\includegraphics[width=8cm,clip]{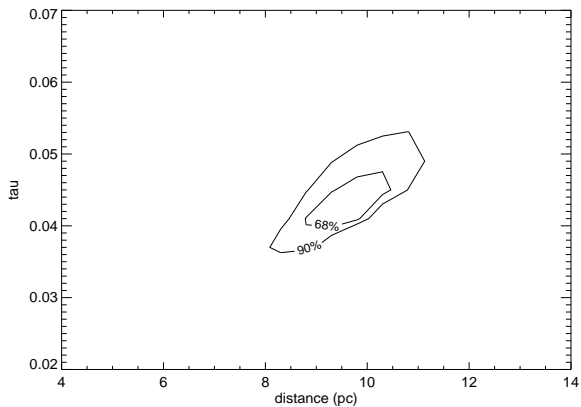}} 
\caption{68\% and 90\% confidence levels of the distance and $\tau$ in the FUV towards the South of Spica.}
\end{figure}

\begin{figure}[h]
\centerline{\includegraphics[width=8cm,clip]{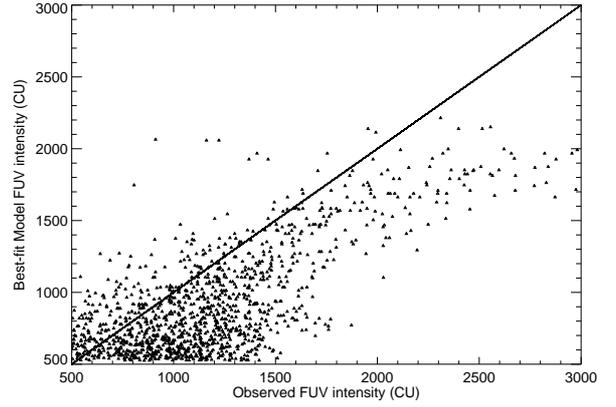}} 
\caption{Comparison of observed and model FUV intensities toward the South of Spica.}
\end{figure}

\begin{figure}[h]
\centerline{\includegraphics[width=8cm,clip]{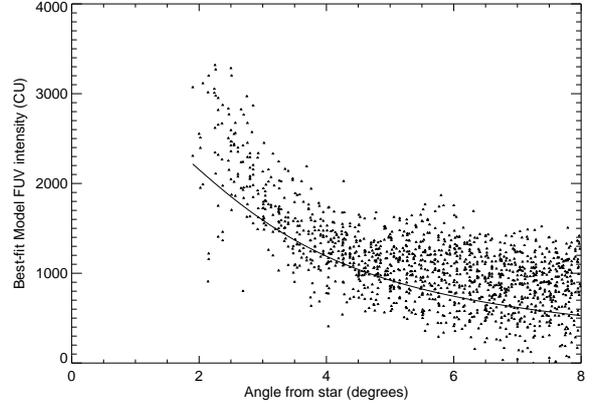}} 
\caption{Angular dependence of the observed and model FUV intensity of the halo towards the South of Spica.}
\end{figure}
For Southern parts of the UV halo, we were unable to uniquely constrain the dust parameters using our model. This is because of the complicated dust distribution in the region, probably due to the presence of multiple dust layers here. However, we have explored one possibility of assuming the dust properties to be similar everywhere (North and South of Spica) and derived the model intensities for the best-fit albedo and $g$ values derived for the Northern regions. Since the extinction is known to be higher for this region (Park et al. 2011) we have used a higher total $\tau$ of 0.2 for the two dust sheets. In this way we could constrain the distance and $\tau$ of the scattering layer to 9.5$\pm$1.5 pc and 0.04$\pm$0.01 from the FUV model (see Figure 5).  
By comparing the observed and best-fit model intensities (Figure 6) we get a correlation co-efficient of 0.75 showing that the model is not as good a fit to the observed intensities as that obtained for the Northern region. 
From the figure, it is also evident that there is a large spread in the observed intensities for this region. 
The plot of the angular dependence of the observed and model intensities (see Figure 7) also shows that the model is not a very good representation of the observed profile.

\section{Conclusions}
Using a simple single scattering model, we have been able to constrain uniquely the optical constants as well as the dust distribution for the Northern parts of the Spica halo. We find that our $g$ values are consistent with previous results of Murthy et al. (2011) as well as with the theoretical predictions of Draine (2003) while our albedos are lower. This may be a consequence of the uncertain dust distribution or else a real difference in the dust properties in the region which was observed by Zagury et al. (1998). In the Southern regions, the complicated dust geometry makes it difficult to uniquely determine the parameters. We have however explored one possibility where the dust has higher densities and placed constraints on the dust geometry in the region. 
However, detailed modeling by dividing the region into smaller zones (due to the large scatter in the observed intensities) may be the only solution to uniquely determining the dust properties to the South of Spica. In addition, the inclusion of NUV and IR models could also greatly contribute in more accurate determinations of these parameters in the region.
%%\lastpagecontrol{90cm}
\acknowledgments{We sincerely thank the organisers of the Cosmic Dust meeting for giving us the opportunity to present our work. We thank the referees for their suggestions and comments which helped us in improving our manuscript. We are also grateful to Dr. H. Kimura for useful suggestions.}

\email{P.Shalima (e-mail: shalima.p@gmail.com)}
\label{finalpage}
\lastpagesettings
\end{document}